\documentclass[twocolumn]{aastex6}

\usepackage{graphicx}                    

\begin{document}

\title{Current Sheet Structures Observed by the TESIS EUV Telescope During A Flux Rope Eruption on the Sun}

\author{A.A.~Reva, A.S.~Ulyanov, and S.V.~Kuzin}
       
\affil{Lebedev Physical Institute, Russian Academy of Sciences;
                     email: \url{reva.antoine@gmail.com} }

\begin{abstract}
We use the TESIS EUV telescope to study the current sheet signatures observed during flux rope eruption. The special feature of the TESIS telescope was its ability to image the solar corona up to a distance of 2~$R_\odot$ from the Sun's center in the Fe~171~\AA\ line. The Fe~171~\AA\ line emission illuminates the magnetic field lines, and the TESIS images reveal the coronal magnetic structure at high altitudes. The analyzed CME had a core with a spiral---flux rope---structure. The spiral shape indicates that the flux rope radius varied along its length. The flux rope had a complex temperature structure: cold legs (70~000~K, observed in He~304~\AA\ line) and a hotter core (0.7~MK, observed in Fe~171~\AA\ line). Such structure contradicts the common assumption that the CME core is a cold prominence. When the CME impulsively accelerated, a dark double Y-structure appeared below the flux rope. The Y-structure timing, location, and morphology agree with the previously performed MHD simulations of the current sheet. We interpreted the Y-structure as a hot envelope of the current sheet and hot reconnection outflows. The Y-structure had a thickness 6.0~Mm. Its length increased over time from 79~Mm to more than 411~Mm.
\end{abstract}
\keywords{Sun: corona---Sun: coronal mass ejections (CMEs)}

\section{Introduction}

Coronal mass ejections (CMEs) are giant eruptions of the coronal plasma into the interplanetary space. CMEs result from processes of energy release in the solar corona. When CMEs reach the Earth, they affect the space weather. The CME investigations are important for solar physics and for the questions of solar-terrestrial connections.

\begin{figure*}[hbtp]
\centering
\includegraphics[width = \textwidth]{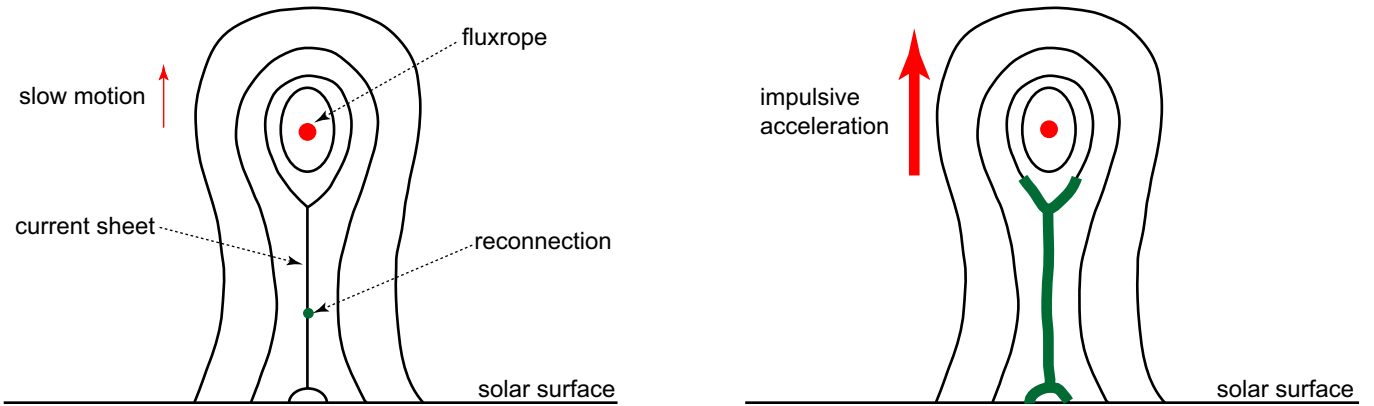}
\caption{Standard CME model. Left: CME before reconnection; right: CME after reconnection. Green indicates the hot plasma in the current sheet and reconnection outflows.}
\label{F:CSHKP}
\end{figure*}

According to the standard CME model \citep{Carmichael1964, Sturrock1966, Hirayama1974, Kopp1976}, before eruption, the CME structure is the flux rope (an elongated twisted magnetic field structure). The flux rope is located above the current sheet and the arcade (see Figure~\ref{F:CSHKP}). Due to various reasons, the flux rope starts to slowly move up. This motion stretches the current sheet, and reconnection occurs. The plasma outflow from the reconnection region accelerates the flux rope and the CME erupts.

The flux rope and the current sheet are essential parts of CME models \citep{Chen2011}. To build a comprehensive picture of solar phenomena, we need to investigate flux ropes and current sheets: measure their physical parameters and find evidence of their existence.

Since we cannot measure the coronal magnetic field, we cannot prove that the flux ropes exist in the corona. Only circumstantial evidence supports their existence: a twisted magnetic field on the photosphere \citep{Lites2005}, presence of the flux ropes in the reconstructed magnetic field \citep{Yan2001, Schrijver2008}, helical structures in the Doppler field maps \citep{Ciaravella2000} and in the EUV telescopes images \citep{Nindos2015}. 

For the same reasons (we can measure neither the magnetic field nor the currents in the corona), it is impossible to directly observe the current sheet in the corona.  Moreover, the current sheet should be a thin structure with low emission, and its observational signatures are expected to be weak. However, several indirect current sheet observations are available: spectroscopic observations of hot lines emission from the current sheet-like structures \citep{Ciaravella2002, Ko2003, Ciaravella2008}, darkening in the Ly$\alpha$ line below the CME \citep{Lin2005}, and elongated linear structure above the flare arcade in the soft-X-ray images \citep{Savage2010}.

The flux rope and the current sheet are one of the most widely accepted and vital concepts in the solar physics. At the same time, they are the hardest to observe. Despite several observations of the flux rope and current sheet signatures, there is a need for more independent evidence of their existence. 

TESIS is a set of the EUV and soft X-ray telescopes intended to investigate the solar corona \citep{Kuzin2011}. A special feature of the TESIS EUV telescopes was the ability to  image the solar corona in the Fe~171~\AA\ line up to the distance of 2~$R_\odot$ from the Sun's center. The Fe~171~\AA\ line illuminates magnetic field lines, and the TESIS images reveal magnetic structures at high altitudes. 

On April 17, 2009, TESIS observed a CME that showed the flux rope and the current sheet signatures. The studied CME had a core with a well-distinguished spiral---flux rope---structure. When the CME impulsively accelerated, the current sheet-like structure appeared below the flux rope. In this work, we will describe the flux rope-like CME core, the current sheet-like structure, and study the connection between the CME acceleration and the current sheet.

\section{Experimental Data}

The analyzed CME occurred on April 17, 2009 at the north-eastern part of the solar limb. For the analysis, we use the data of the TESIS EUV telescopes \citep{Kuzin2011},  LASCO coronographs \citep{Brueckner1995}, the {\it Extreme ultraviolet Imaging Telescope} \citep[EIT;][]{del95}, and the data of the Sphinx spectrophotometer \citep{Gburek2011}.

TESIS  is an instrument assembly that investigated the solar corona in the EUV and soft X-ray wavelength ranges. TESIS worked on board \textit{CORONAS-PHOTON} satellite \citep{Kotov2011}. TESIS included EUV telescopes, which built images of solar corona in Fe~171~\AA\ and He~304~\AA\ lines with $1.7^{\prime\prime}$ spatial resolution. A special feature of the Fe~171~\AA\ telescope was its ability to image the solar corona up to 2~$R_\odot$ \citep[for more details see][]{Reva2014}. In the `far corona' mode, TESIS registered `binning' images with the $3.4^{\prime\prime}$ resolution. During the period of the observations, TESIS had a varying cadence of 30~min and 1~hr. The TESIS data are central in our research---we use them to study coronal structures at high altitudes.

TESIS `far corona' images are composed out of three images with different exposure times: short, medium, and long. Since all three images are taken with the same instrument, the result looks like a genuine single-shot image. However, the merging algorithm is not perfect, and on some of the images, the border between images with different exposures can be seen.

LASCO is a set of white-light coronagraphs, which observe solar corona from 1.1~$R_\odot$ up to 30~$R_\odot$: C1: 1.1--3~$R_\odot$, C2: 2--6~$R_\odot$, C3: 4--30~$R_\odot$. In 1998, LASCO C1 stopped working, and today LASCO can only image corona above 2~$R_\odot$. We use LASCO data to complement the TESIS data above 2~$R_\odot$.

The EIT on SOHO satellite takes solar images at the wavelengths centered at 171, 195, 284, and 304~\AA. In a synoptic mode, EIT takes images in all four channels every 6 h; in the `CME watch' mode, the telescope takes images in the 195~\AA\ channel every 12~min. The pixel size of the telescope is $2.6^{\prime\prime}$, and the spatial resolution is $5^{\prime\prime}$. We use the EIT data to complement TESIS data with the observations in the 195 and 284~\AA\ lines.

Sphinx is a spectrophotometer that studied the Sun in soft X-ray. It worked on board \textit{CORONAS-PHOTON} satellite. Sphinx registered solar spectra in the 1--15~keV energy range. In 2009, the solar cycle was in deep minimum, and the GOES flux usually was below the sensitivity threshold. Sphinx is more sensitive than GOES, and we use it to see the variation of the X-ray flux.

\section{Results}

\subsection{CME Evolution}

\begin{figure*}[t]
\centering
\includegraphics[width = \textwidth]{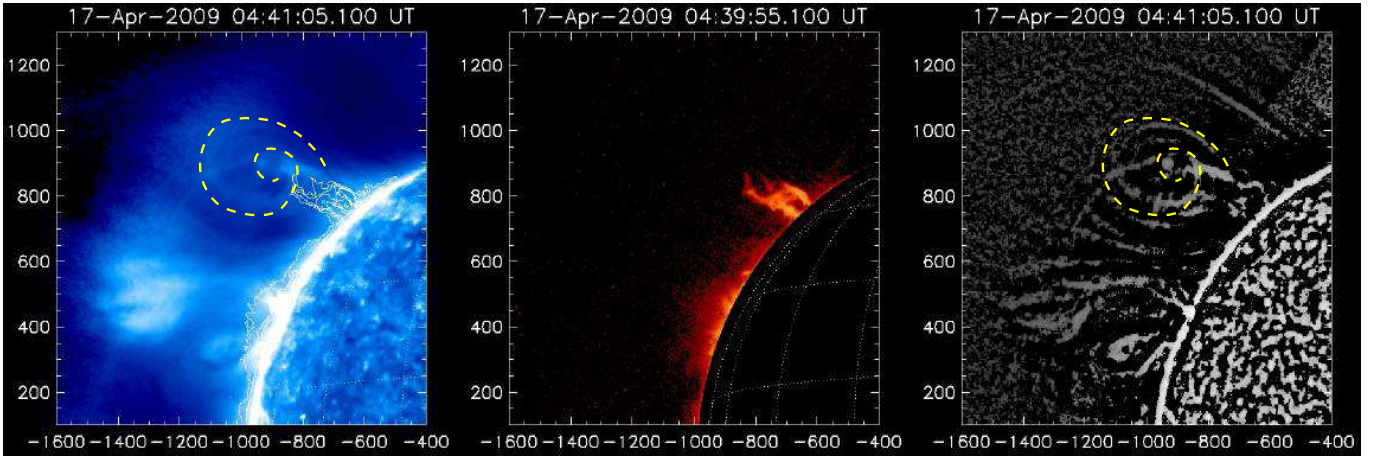}
\caption{Left: TESIS Fe~171~\AA\ image; middle: TESIS He~304~\AA\ image; right: sharpened TESIS Fe~171~\AA\ image. To improve the prominence visibility, we applied an artificial oculting disk to the He~304~\AA\ image. Yellow dashed line marks the spiral structure. Coordinates are measured in arc seconds.}
\label{F:Fe_he_spiral}
\end{figure*}

\begin{figure*}
\centering
\includegraphics[width = \textwidth]{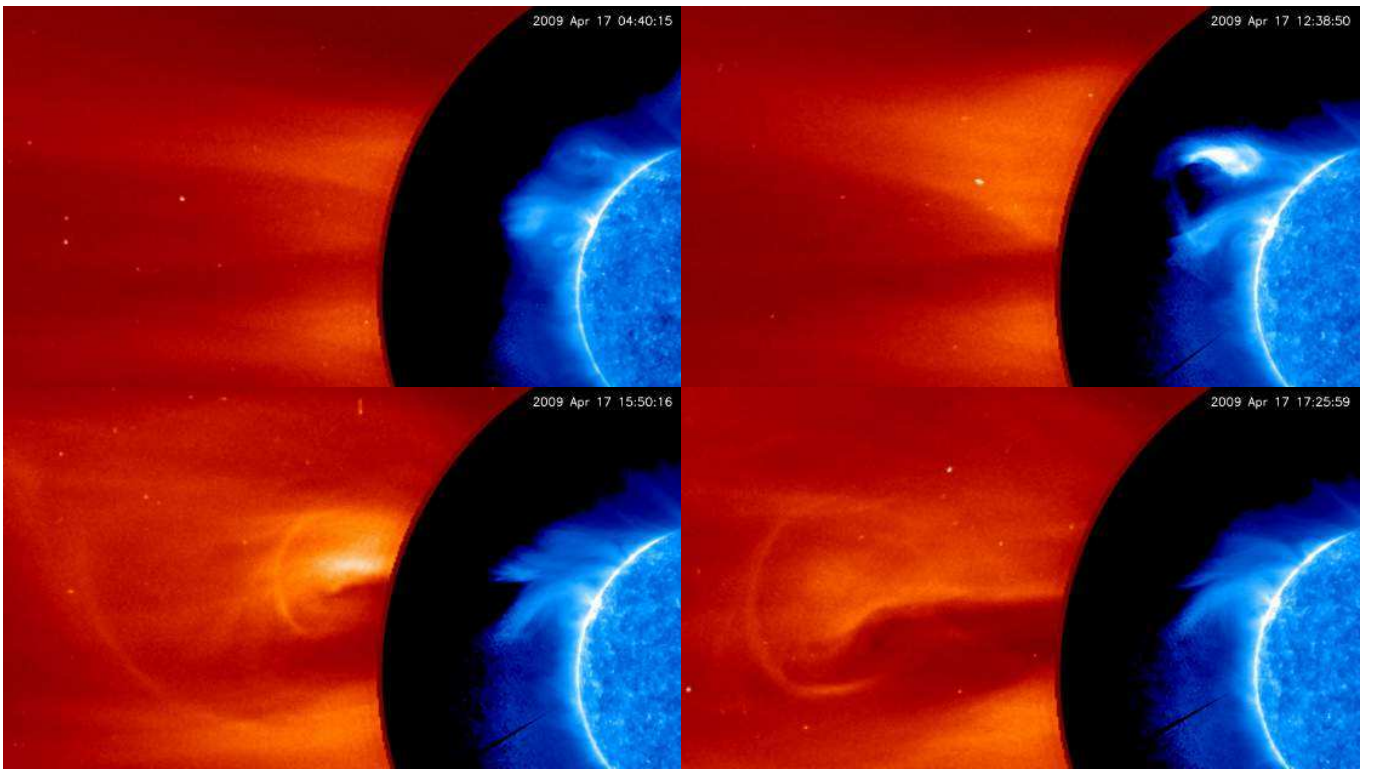}
\caption{CME propagation. Blue: TESIS Fe~171~\AA\ telescope; red: LASCO C2. An animation of this figure is available.}
\label{F:Fe_c2}
\end{figure*}

On April 16, 2009 at 02:32~UT, a spiral structure---flux rope---formed on the TESIS Fe~171~\AA\ images. In the He~304~\AA\ images, there was a prominence below the flux rope, and there was no cool prominence plasma in the flux rope center (see~Figure~\ref{F:Fe_he_spiral}). 

At 05:32~UT on April~16, the flux rope moved up and expanded its diameter. While moving up, the flux rope preserved its spiral shape. At 09:57~UT on April~17, a darkening appeared below the flux rope. The darkening expanded and took the shape of the Y-structure. 

When the CME reached the LASCO/C2 field of view, it had a 3-part structure: a bright frontal loop, a dark cavity, and a bright core (see~Figure~\ref{F:Fe_c2}). An animation of Figure~\ref{F:Fe_c2} shows how the analyzed event evolved. We recommend the reader to watch this video.

\subsection{Kinematics}

\begin{figure*}[t]
\centering
\includegraphics[width = \textwidth]{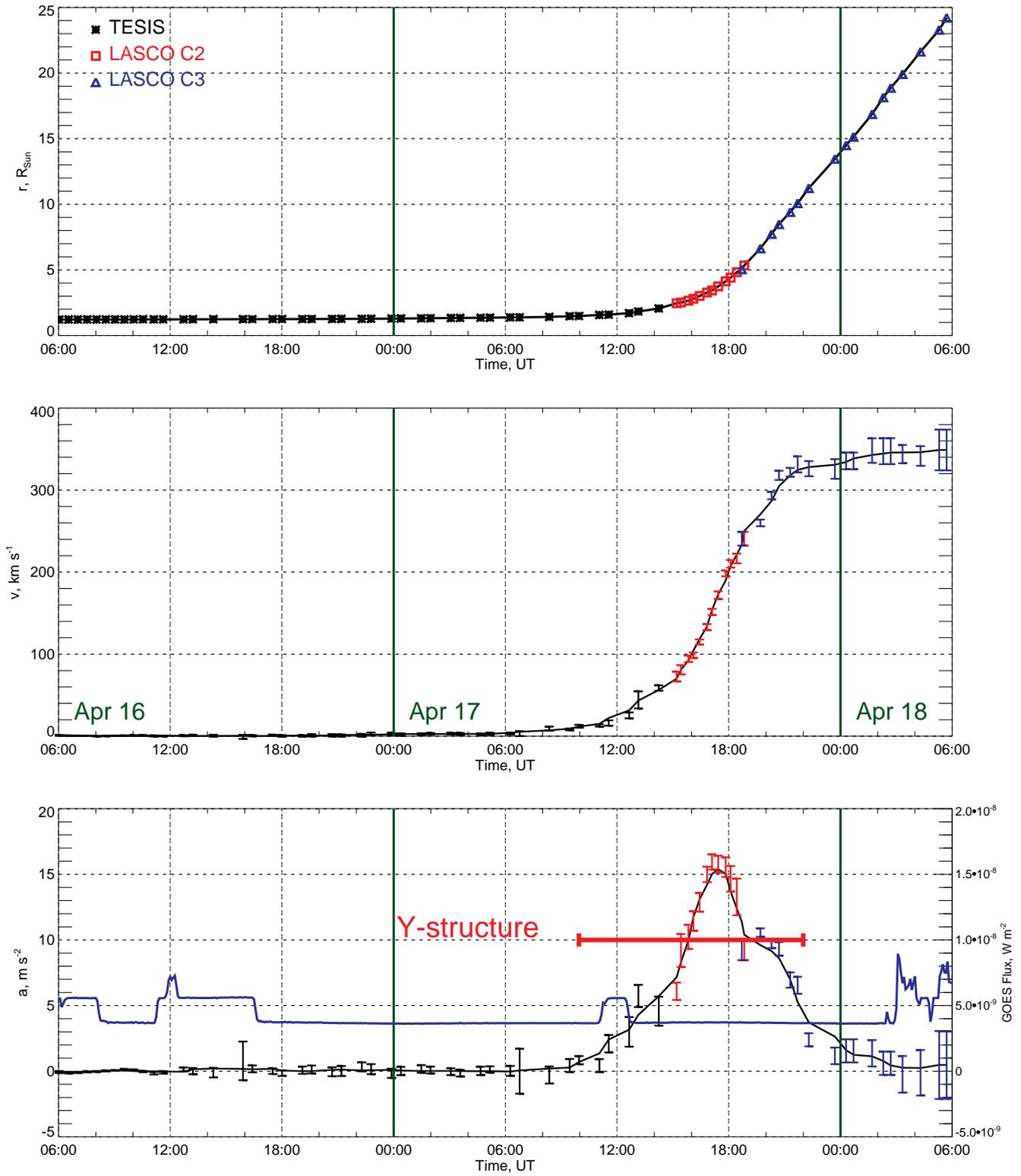}
\caption{Flux rope kinematics. $r$---distance from the flux rope to Sun's center; $v$---flux rope velocity; $a$---flux rope acceleration. Black asterisks: TESIS Fe~171~\AA\ data; red squares: LASCO C2 data; blue triangles: LASCO C3 data. Blue line indicates flux in the GOES 0.5--4~\AA\ channel.}
\label{F:Kinematics}
\end{figure*}

We measured coordinates of the flux rope center in the TESIS and LASCO images. The measurements were carried out with a simple point-and-click procedure. To estimate error bars, we repeated the procedure 9 times. We numerically differentiated the flux rope radial coordinate $r(t)$ and obtained its radial velocity $v(t)$ and radial acceleration $a(t)$ (see Figure~\ref{F:Kinematics}).

\begin{figure*}[t]
\centering
\includegraphics[width = \textwidth]{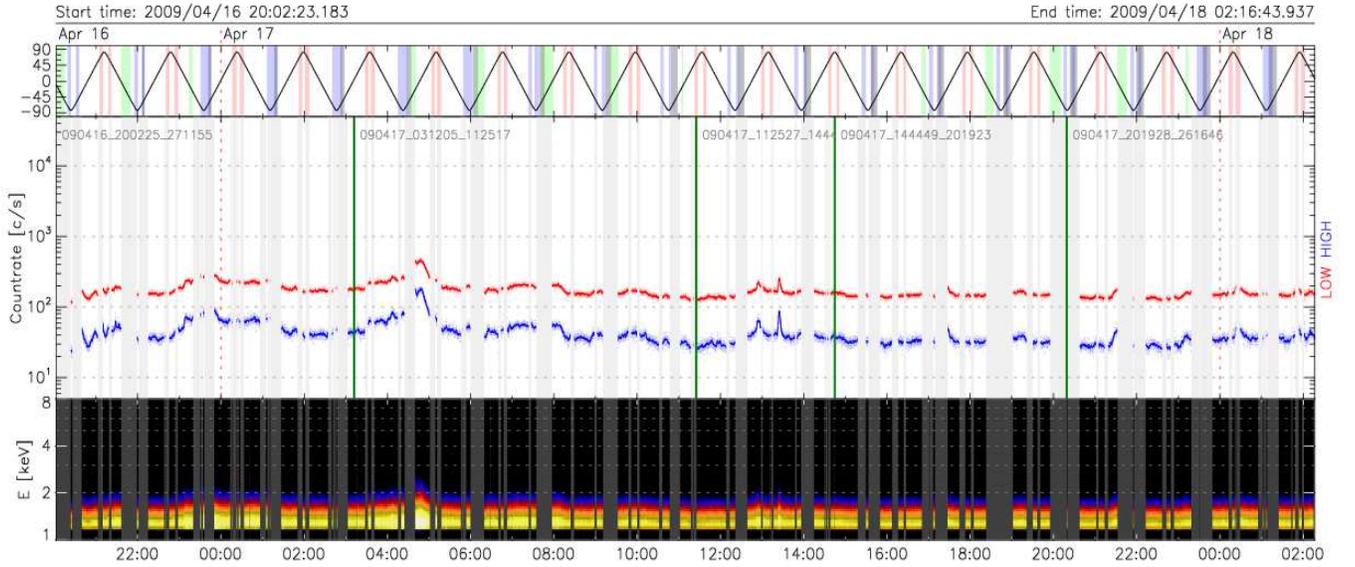}
\caption{Sphinx countrates. Red: 1.2--1.5~keV energy range, blue: 1.5--15.0~keV energy range. Bottom panel: a dynamic Sphinx spectrum. Top panel: a spacecraft orbit latitude. The image was taken from \url{http://www.cbk.pan.wroc.pl/?l=EN&act=6}.}
\label{F:Sphinx}
\end{figure*}

The CME kinematics consisted of the three phases: slow motion, impulsive acceleration, and propagation with constant velocity. During the slow motion phase, the flux rope moved with a velocity of 0.5--3~km~s$^{-1}$. During the impulsive acceleration phase, from 10:00~UT April 17 to 00:00~UT April 18, the flux rope accelerated up to 350~km~s$^{-1}$ with a peak acceleration of 15~m~s$^{-2}$. After 00:00~UT April 18, the flux rope moved with a constant velocity.

The studied CME has average kinematics properties. Its slow motion velocity coincides with the slow motion velocity of the erupting prominencies \citep[0.5--4~km~s$^{-1}$,][]{McCauley2015}. Its final speed is close to the average CME speed during the solar minimum \citep[280~km~s$^{-1}$,][]{Yashiro2004}. Its acceleration lies between -10 and +20~m~s$^{-2}$---typical values for CMEs with speed between 250 and 450~km~s$^{-1}$ \citep{Yashiro2004}.

During the observation period, the GOES 1--8.0~\AA\ flux did not exceed the sensitivity threshold, and the 0.5-4~\AA\ flux was below the A-level (see Figure~\ref{F:Kinematics}). The GOES signal did not correlate with the CME acceleration. The Sphinx registered two microflares at the beginning of the CME impulsive acceleration phase (at 12:55~UT and 13:25~UT, see Figure~\ref{F:Sphinx}).

\subsection{Y-structure} 

\begin{figure*}[hbt]
\centering
\includegraphics[width = 0.95\textwidth]{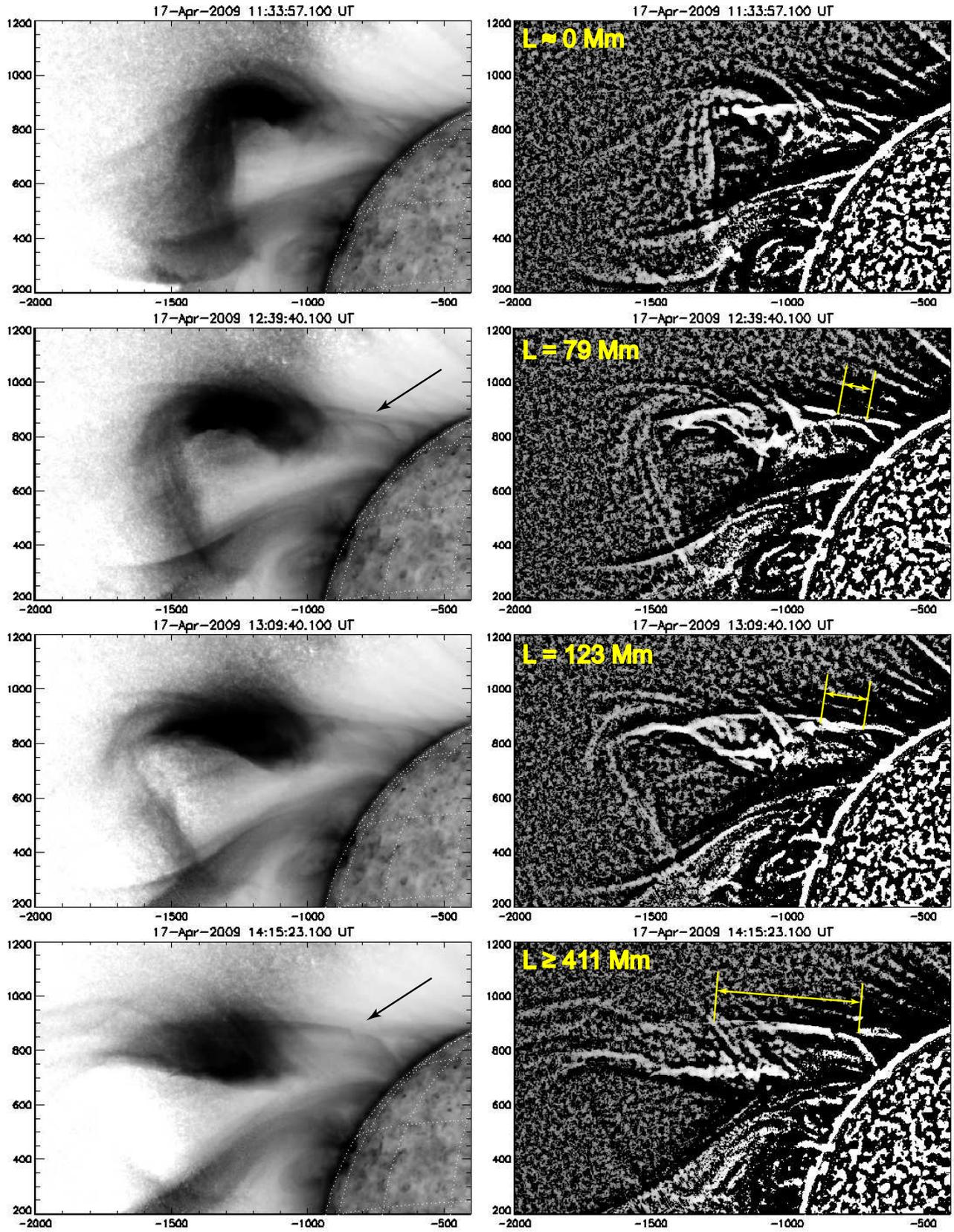}
\caption{Y-structure formation. Left: negative images (dark corresponds to high intensities, bright to low). Arrows indicate the Y-structure. Right: sharpened images. Yellow lines denote current sheet boundaries. Coordinates are measured in arc seconds.}
\label{F:Y_structure}
\end{figure*}

\begin{figure*}[t]
\centering
\includegraphics[width = 0.49\textwidth]{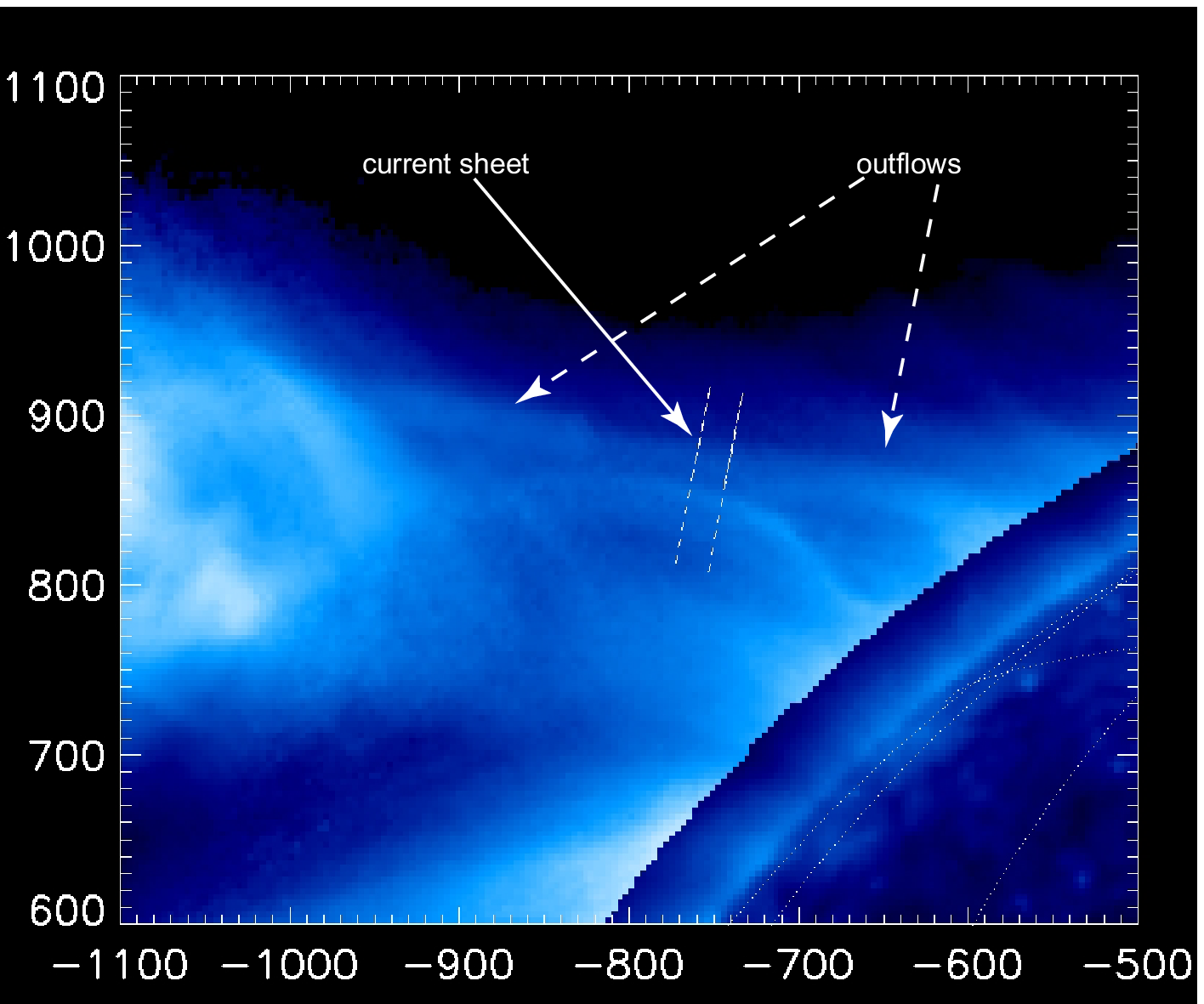}
\includegraphics[width = 0.49\textwidth]{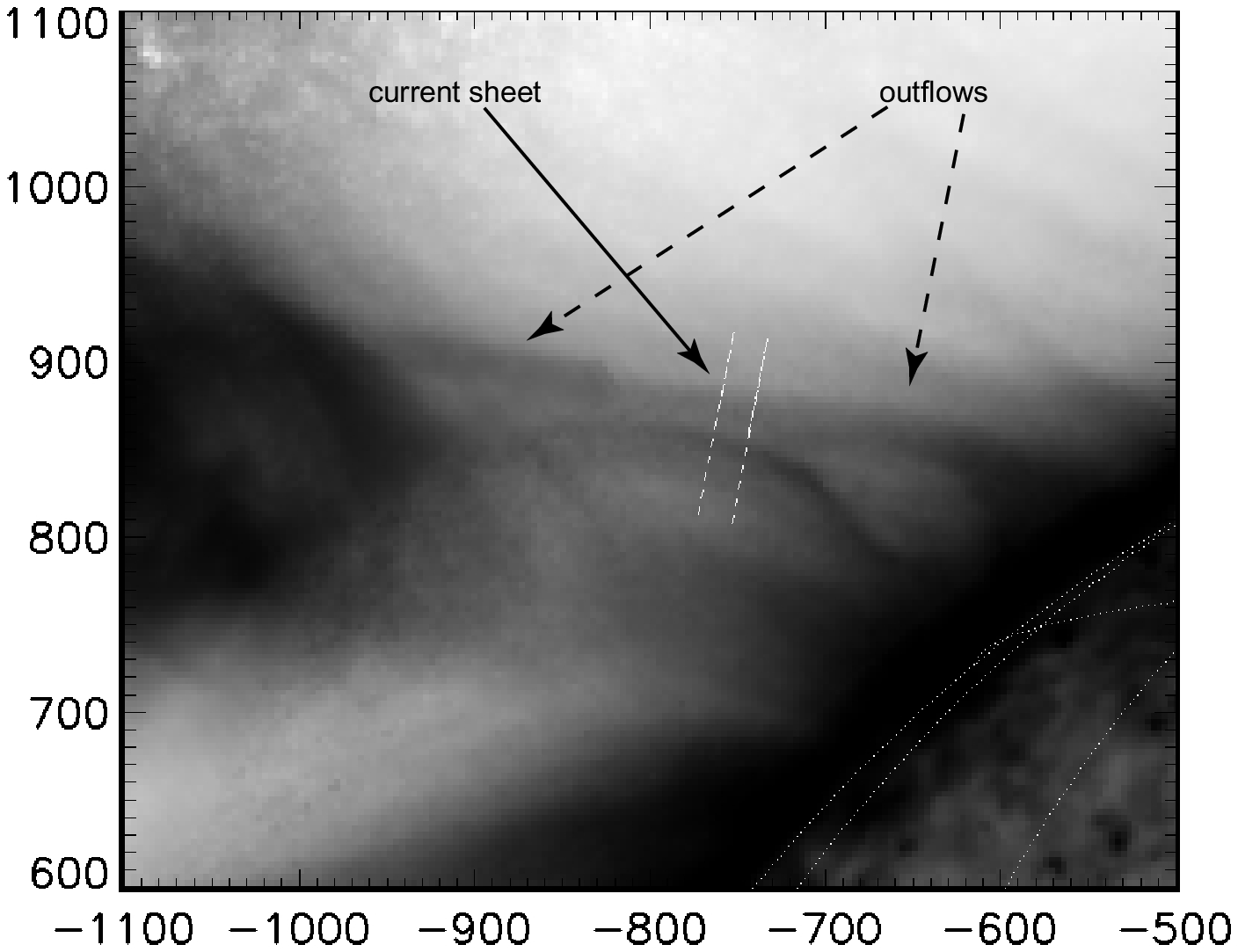}
\caption{Current sheet and reconnection outflows. Left: normal image (bright corresponds to high intensities, dark to low). We artificially lowered the solar disk intensity to make the current sheet more distinguishable. Right: negative image (dark corresponds to high intensities, bright to low). Image was taken on April 17 at 12:38~UT. The dashed lines designate the position of the artificial slit. Coordinates are measured in arc seconds.}
\label{F:Y_outflows}
\end{figure*}

Under the flux rope, 60~Mm above the solar surface at 09:57~UT on April~17, the small darkening appeared (see Figure~\ref{F:Y_structure}). The darkening expanded and took the shape of the double Y-structure. Approximately at 22:00~UT, the Y-structure disappeared. The Y-structure appeared at the same time as the CME impulsively accelerated (see Figure~\ref{F:Kinematics}). The shape of the Y-structure---an elongated linear structure and widenings at its ends---highly resembles a current sheet (see Figure~\ref{F:Y_outflows}). 

In the Fe~171~\AA\ line, a dark Y-structure appeared because either the density decreased, the temperature decreased, or the temperature increased. To distinguish between these cases, we compared the TESIS images of the Y-structure with the EIT 195 and 284~\AA\ images.

\begin{figure*}[hbt]
\centering
\includegraphics[width = \textwidth]{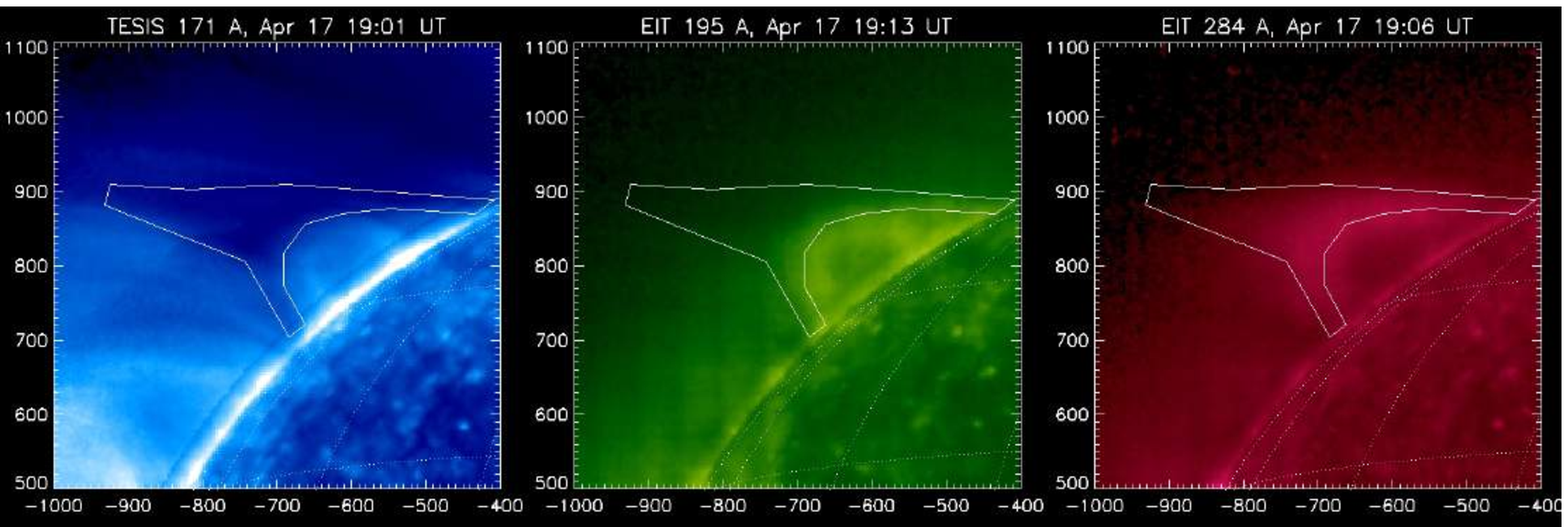}
\caption{Flare loop. Left: TESIS 171~\AA\ image; middle: EIT 195~\AA\ image; right: EIT 284~\AA\ image. Contours denotes dark Y-structure, which is present on the TESIS 171~\AA\ images.}
\label{F:flare_loop}
\end{figure*}

\begin{figure}[hbt]
\centering
\includegraphics[width = 0.45\textwidth]{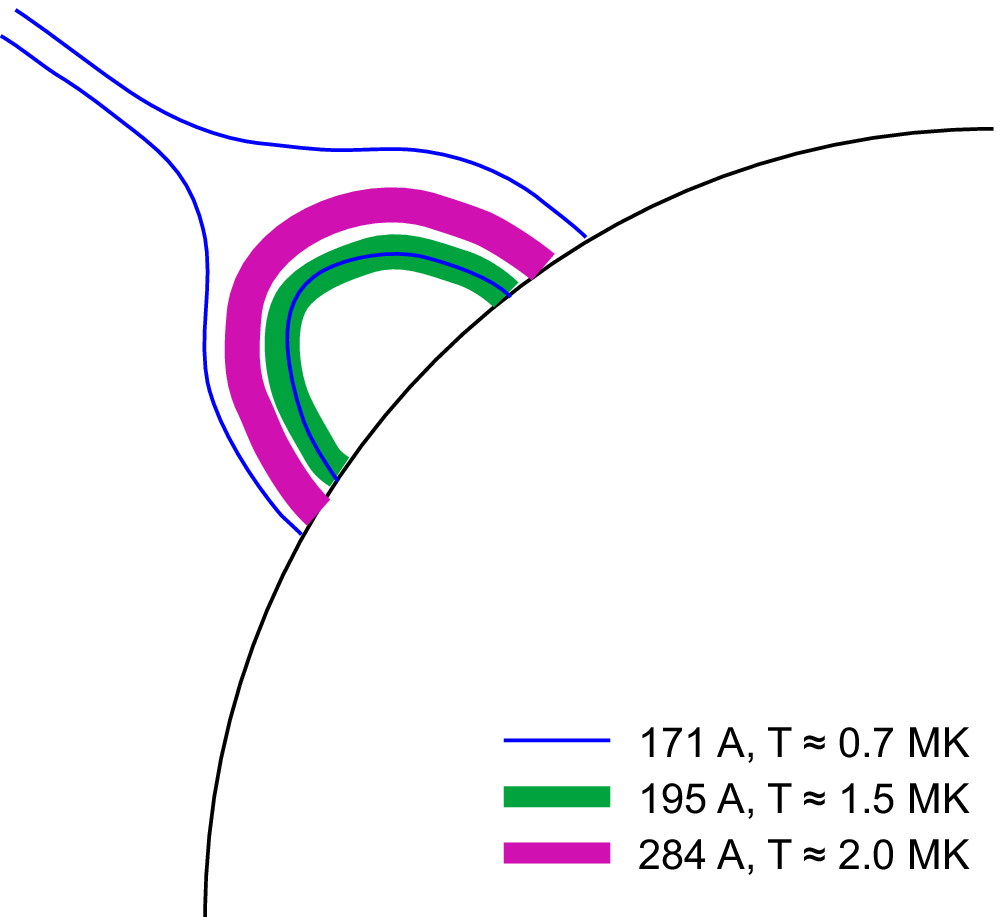}
\caption{Relative positions of the dark Y-structure observed with the TESIS Fe 171~\AA\ telescope (blue) and flare loops observed with the EIT 195~\AA\ (green) and 284~\AA\ (purple).}
\label{F:tesis_eit_sketch}
\end{figure}

Figure~\ref{F:flare_loop} shows a flare loop, which formed under the flux rope. Contours denote the dark Y-structure. Half of the loop in the EIT 195~\AA\ and almost the entire loop in the EIT 284~\AA\ images lie within the Y-structure contour (see Figures~\ref{F:flare_loop},~\ref{F:tesis_eit_sketch}). Since the flare loops in the EIT 195 and 284~\AA\ images are located inside the Y-structure, then the Y-structure was not a density deficit, but rather a temperature change. The 195~\AA\ line emits at temperatures around 1.5~MK, and the 284~\AA\ line emits at temperatures around 2~MK. Therefore, the dark Y-structure was a hot plasma, but not a cold plasma or a density deficit.

\subsection{Y-structure Thickness}

\begin{figure*}[hbt]
\centering
\includegraphics[width = 0.49\textwidth]{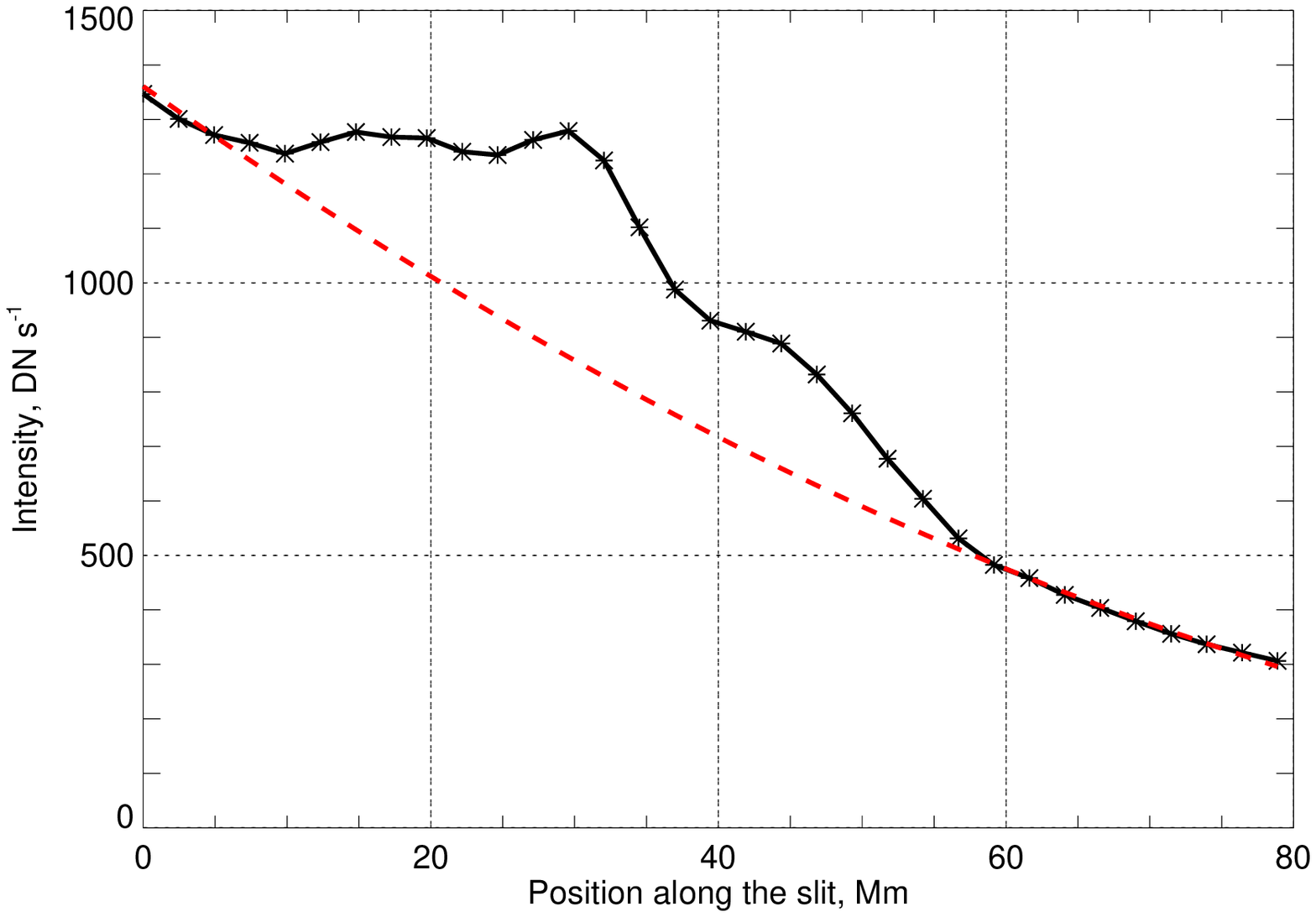}
\includegraphics[width = 0.49\textwidth]{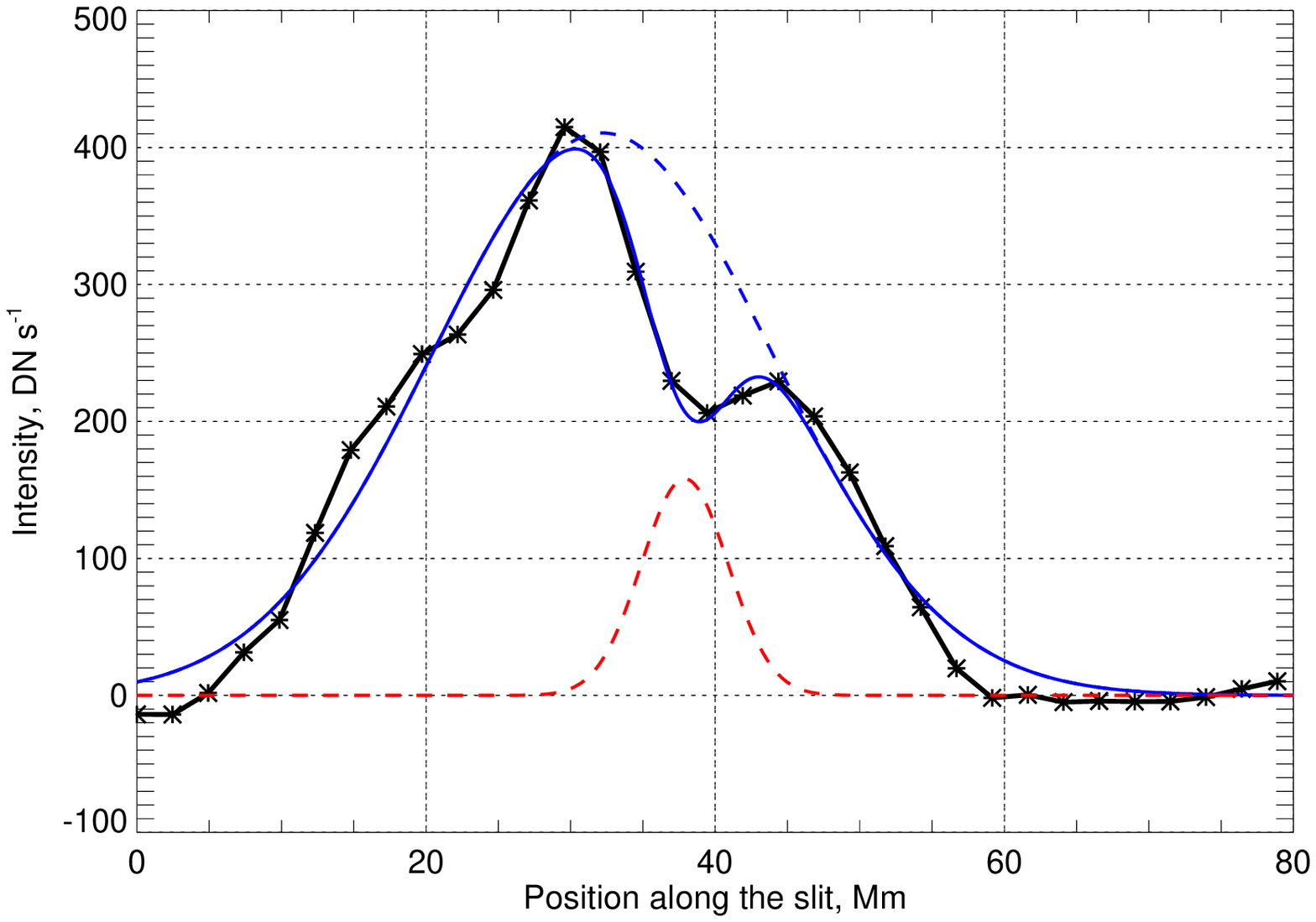}
\caption{Left: scanning of the Y-structure intensity. Black line indicates the intensity across the slit, red dashed line indicates background fit. Right: determination of the Y-structure thickness. Black line: intensity scan after background subtraction; dotted blue line: `main' Gaussian profile; dotted red line: Y-structure Gaussian profile; blue line: intensity fit (difference between the `main' and the Y-structure Gaussian profiles).}
\label{F:cs_scan}
\end{figure*}

To determine the Y-structure thickness, we cut a slice from the TESIS image in the direction perpendicular to the Y-structure (see Figure~\ref{F:Y_outflows}). The length of the slice is 80~Mm, and the width is 15~Mm. We summed the slice intensity along its width (summed the pixels perpendicular to and in between the two white lines in the Figure~\ref{F:Y_outflows}) and obtained the intensity scan (see Figure~\ref{F:cs_scan}, left).

The intensity scan consists of a signal from the `bright envelope' (bright areas around the dark Y-structure), the dip in the intensity (the Y-structure), and the signal from the background corona. We fitted the background with a second order polynomial and subtracted it from the intensity scan (see Figure~\ref{F:cs_scan}, right).

The remaining signal consists of the `bright envelope' and the intensity dip. We fitted the remaining signal with the formula:

\begin{equation}
I(x) = I_1 \exp \left(-\frac{(x-x_1)^2}{2\sigma_1^2}\right) - I_2 \exp\left(-\frac{(x-x_2)^2}{2\sigma_2^2}\right),
\end{equation}

where $I_1$---the `bright envelope' intensity, $x_1$---the position of the envelope maximum, $\sigma_1$---the width of the envelope, $I_2$---the Y-structure `intensity', $x_2$---the position of the Y-structure, and $\sigma_2$---the width of the Y-structure. After fitting, we obtained that $\sigma_2$~=~3.0~Mm. We estimated the Y-structure thickness as $h = 2\sigma_2 =$~6.0~Mm (1.8~pixels).

\subsection{Y-structure Length}

The Y-structure length---the distance between upper and lower widenings---increased as the CME moved up (see Figure~\ref{F:Y_structure}). At 12:39 and 13:09~UT, we can reliably measure its length directly from the images (79~Mm at 12:39~UT and 123~Mm at 13:09~UT).

At 11:33~UT, we see the dark downward outflow, but do not see the Y-structure. We think that this is because the Y-structure length was so small that TESIS could not resolve it. We estimate the Y-structure length at 11:33~UT to be almost zero.

At 14:15~UT, the Y-structure was too large, and TESIS saw only a part of it. Its length---411~Mm---is a lower estimate of the Y-structure length.

\begin{figure}[hbt]
\centering
\includegraphics[width = 0.45\textwidth]{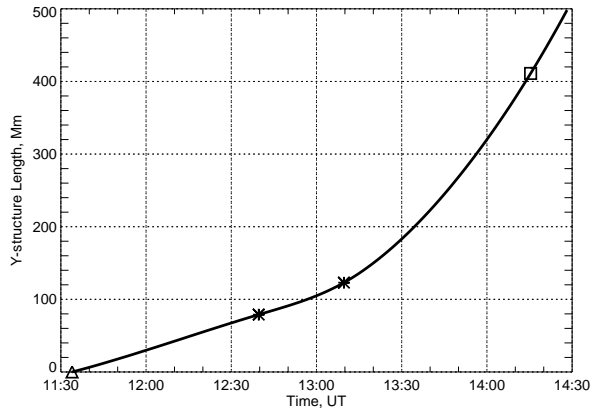}
\caption{Y-structure length as a function of time. Asterisks denote directly measured values. Triangle denote very small length that we estimated as zero. Square denotes the lower estimate of the Y-structure length.}
\label{F:cs_length}
\end{figure}

We put these values on a single plot (see Figure~\ref{F:cs_length}). We must stress that this is a very rough plot: only half of the points are real measurements; the other half are estimates. The plot illustrates how the Y-structure length increased over time.

\section{Discussion}

We presented observations of the eruptive flux rope made with the TESIS EUV telescopes in the Fe 171~\AA\ and He~304~\AA\ lines. The flux rope had spiral structure with a complex temperature structure: cool legs and a hotter core. When the CME impulsively accelerated, the dark double Y-structure formed below the flux rope in the Fe 171~\AA\ images. The double Y-structure highly resembles structures that are predicted to form around the current sheet. Below we will discuss the Y-structure and the flux rope in further detail.

\subsection{Current Sheet}

During reconnection, around the current sheet an envelope of hot plasma should form \citep{Yokoyama1998, Seaton2009, Reeves2010}. This effect broadens the visible thickness of the current sheet. \citet{Reeves2011} performed MHD simulation of the current sheet and produced synthetic AIA 171 \AA\ images of the simulated current sheet. In the synthetic images, the current sheet looked like a dark Y-structure surrounded by a bright envelope (see Figure~5 in \citet{Reeves2011}). 

According to the standard model and the MHD simulations, we should observe in the Fe~171~\AA\ line a dark double Y-structure surrounded by the bright envelope. The Y-structure should appear below the flux rope and above the flare arcade during the CME impulsive acceleration.

In the observations, we see that the Y-structure:
\begin{enumerate}
\item Had a double Y-shape;
\item Was dark in the TESIS Fe~171~\AA\ images;
\item Occurred below the flux rope and above the flare arcade;
\item Occurred during the CME impulsive acceleration;
\item The bottom part of the Y-structure was filled with hot plasma imaged by EIT 284~\AA;
\item Two microflares occured at the beginning of the CME impulsive acceleration phase.
\end{enumerate}

The observations agree with the predictions of the standard model and the MHD simulations. We interpret the Y-structure as a hot envelope of the current sheet and the hot reconnection outflows. The observed thickness of the Y-structure is a thickness of the hot envelope. The real thickness of the current sheet should be lower than the Y-structure thickness. However, these observations are not a 100\% proof. The Y-structure could be some loop rearrangement or heating that randomly coincided with the eruption. The presented observations are indirect evidence of the current sheet.

\subsection{CME acceleration and flares}

In the standard model, a flare and a CME impulsive acceleration should occur simultaneously. This statement is supported by the observations of the correlation between the CME impulsive acceleration and various flare signatures: GOES flares \citep{Zhang2001}, increase of the hard X-ray flux \citep{Gallagher2003}, the arcades footpoints motion \citep{Qiu2004}, and appearance of the flare arcade \citep{Reva2016}.

The analyzed CME was not associated with any GOES flare. However, the dark double Y-structure indicates that the flare did occur. It was a very small flare that was unnoticed by the GOES. This observations show that even if we did not register a flare during the CME, the CME could still be associated with a very weak undetected flare.

At the beginning of the CME impulsive acceleration phase, Sphinx registered two microflares. In 2009, the solar cycle  was in deep minimum. There were very little activity on the Sun, and it is very likely that these microflares were associated with the Y-structure. However, we cannot determine their location. It is possible, that they occurred at some unrelated to the CME area on the Sun.

\subsection{Flux Rope Geometry}

\begin{figure*}[hbt]
\centering
\includegraphics[width = 0.85\textwidth]{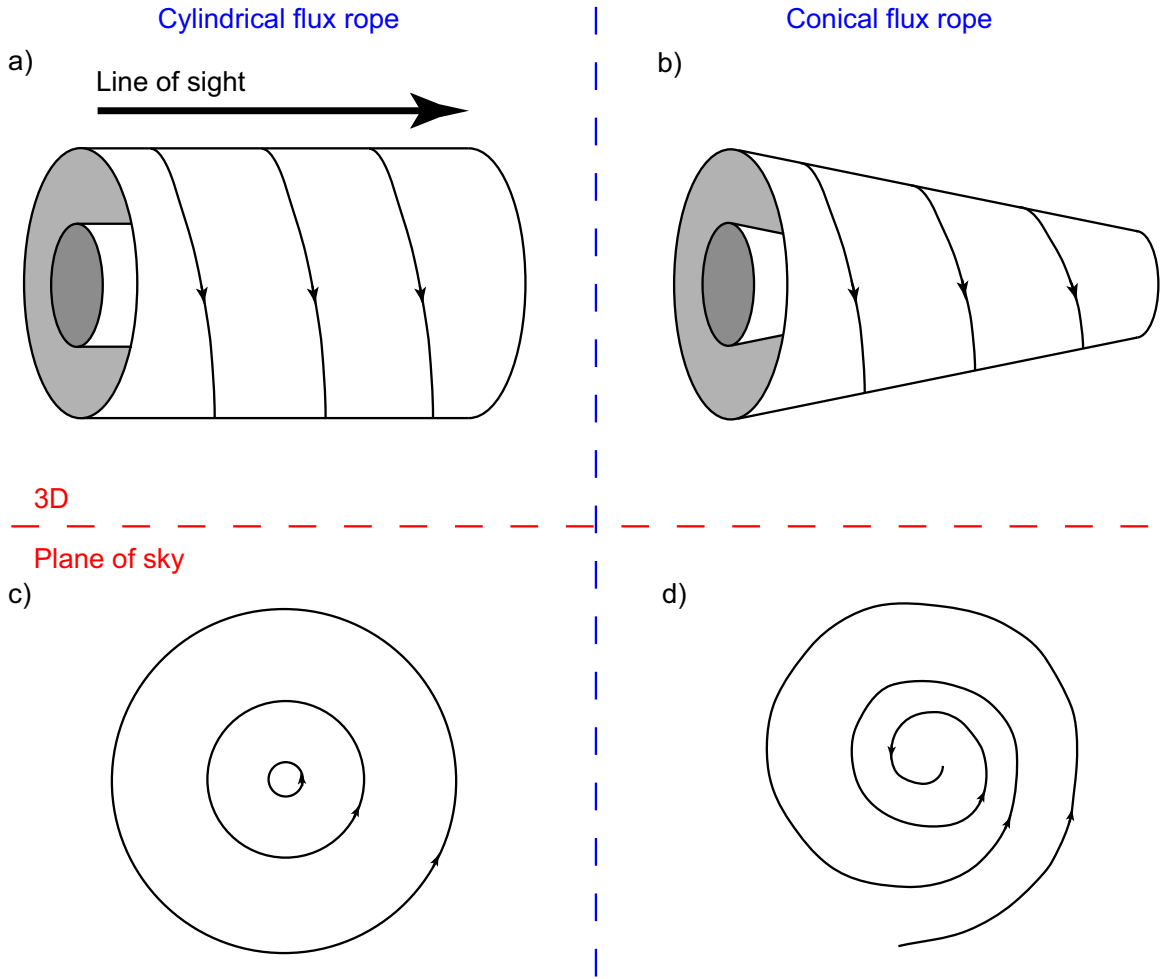}
\caption{Left: cylindrical flux rope. Right: conical flux rope.}
\label{F:Cylinder_cone}
\end{figure*}

The observed CME had a spiral structure: inner parts of the flux rope were connected with its outer parts (see Figure~\ref{F:Cylinder_cone}). In 2.5D, the spiral magnetic field is not divergence free and, therefore, could not exist. The observed spiral structure is a projection of the more complex 3D topology on the TESIS image plane.  We think that the spiral structure indicates that the flux rope diameter varied along its length (see Figure~\ref{F:Cylinder_cone}). 

However, we conclude that the flux rope had spiral shape only based on the TESIS images. TESIS images are a projection of the 3D flux rope structure on the plane of sky. It is possible that unrelated parts of the flux rope superimposed on each other and formed a spiral in the TESIS images.
\subsection{Flux Rope Temperature}

It is generally accepted that a CME core is a prominence, which is located in the flux rope center. Since prominences have a low temperature (70~000~K), it is usually assumed that the CME core also has low temperature. This assumption is supported by several spectroscopic observations  \citep{Akmal2001, Ciaravella1997, Ciaravella1999, Ciaravella2000}, which showed that CME cores had temperatures around 30~000--300~000~K. 

However, in our work, the flux rope was observed both in the He~304~\AA\ line as a prominence, and in the Fe~171~\AA\ line as a spiral structure. The He~304~\AA\ line emits at 70~000~K, and Fe~171~\AA\ line at 0.7~MK. This means that the CME flux rope had cold (70~000~K) prominence legs and a hotter (0.7~MK) flux rope core on the top. Furthermore, we see the passage of the flux rope from the TESIS to the LASCO/C2 field of view almost without losing sight of it. The CME core in the LASCO/C2 images and the flux rope in the TESIS images are the same objects. Therefore, the CME core had temperature around 0.7~MK.

A similar result was obtained earlier.  \citet{Ciaravella2003, Landi2010} showed with spectroscopic measurements that the top of the CME core was hot ($\approx$~1~MK), while its tail was cool ($\approx$~0.1~MK). Likewise, prominences often switch from absorption to emission in the EUV images \citep{Filippov2002}. This effect could indicate heating of the prominence during eruption.

Sometimes, even hotter plasma (5--10~MK) is observed inside the CME core. \citet{Reeves2011} observed a hot plasma inside the CME core in the AIA images. \citet{Song2014} observed 10~MK hot blob in the CME core in the AIA 131~\AA\ images. \citet{Nindos2015} studied limb solar flares (X and M classes) and concluded that half of the CMEs contained hot flux ropes ($T \approx 10$~MK).

We see that the widely spread assumption---that the CME core is a cold prominence---is wrong. The temperature structure of the CME core is a question that requires further investigations.

\section{Conclusion}

In this work, we studied the current sheet and flux rope signatures. For this purpose, we used the data of the TESIS EUV telescopes, because they show the coronal magnetic structure at high altitudes. We chose a CME, for which the magnetic structure of its core was clearly seen. 

The studied CME  had a spiral magnetic structure (flux rope). The flux rope had a complex temperature structure: the cool prominence plasma below the hotter spiral structure. During the CME impulsive phase, a dark double Y-structure (current sheet) formed below the flux rope. The main results of our work are observations of the flux rope geometry, flux rope thermal structure, and the current sheet formation.

The observed spiral structure of the CME is independent evidence that flux ropes exist in the corona. The spiral magnetic field is not divergence-free in 2.5D, and therefore, the flux ropes are essentially 3D objects. We think that the observed spiral structure indicates that the flux rope diameter varied along its length.

It is usually assumed that the CME core is a prominence, and therefore, is cold. However, our and several previous studies have shown that CME cores can have high temperatures. The widespread assumption that the CME core is a cold prominence is wrong. The question about CME core temperature cannot be answered apriori, and it requires further investigation.

The Y-structure timing, location, and morphology agree with the previously performed MHD simulations of the current sheet \citep{Reeves2011}. We interpreted the Y-structure as a hot envelope of the current sheet and the hot reconnection outflows.

In this work, we presented evidence that flux ropes and current sheets exist in the corona. These observations strengthen the experimental foundation of the current sheet and the flux rope concepts.

\acknowledgments
This work was  supported by a grant from the Russian Foundation of Basic Research (grant 14-02-00945).

\bibliographystyle{aasjournal}
\bibliography{mybibl}

\end{document}